\documentclass[12pt]{article}

\usepackage{amsmath}
\usepackage{amsfonts}
\usepackage{amssymb}
\usepackage{pstcol}
\usepackage{pst-plot}
\usepackage[all]{xy}
\usepackage[dvips]{epsfig}

\begin{document}

\title{{\Large{\bf Quantum-gravity-motivated Lorentz-symmetry tests\\
with laser interferometers\\
$~~~$}}
}

\author{{\bf Giovanni Amelino-Camelia}${}^1$
and {\bf Claus L{\"{a}}mmerzahl}${}^2$ \\
$~$\\
${}^1$ Dipart.~Fisica,
Univ.~Roma ``La Sapienza'', and INFN Sez.~Roma1\\
P.le Moro 2, 00185 Roma, Italy\\
amelino@roma1.infn.it\\
$^2$ ZARM, University Bremen, Am Fallturm,
28359, Germany\\
laemmerzahl@zarm.uni-bremen.de}

\maketitle

\begin{abstract}
We consider the implications  for laser interferometry of the
quantum-gravity-motivated modifications in the laws of particle
propagation, which are presently being considered in attempts to
explain puzzling observations of ultra-high-energy
cosmic rays.
We show that there are interferometric setups in which
the Planck-scale effect on propagation leads to a characteristic signature.
A naive estimate is encouraging
with respect to the possibility of achieving Planck-scale sensitivity,
but we also point out some severe technological challenges which would
have to be overcome in order to achieve this sensitivity.
\end{abstract}

\section{Introduction}

Lorentz symmetry plays a key role in our present description
of the fundamental laws of physics, and, as a result, there is
a tradition~\cite{rob,mans1}
of interest in testing this symmetry to the highest possible precision.
During the last few years
the sensitivity of laboratory tests of the principles underlying
Lorentz symmetry has improved
significantly~\cite{Braxmaieretal01,Wolfetal02,Lipaetal03,StAndrewsproceedings}
and this has energized efforts aimed at improving these tests
even further~\cite{Laemmerzahletal01,Buchmanetal00}.
In addition, there has also been strong
interest~\cite{domokos,Amelino-Cameliaetal98,Billeretal99,Schaefer99,glast}
in precision tests of Lorentz symmetry that are based on
certain types of observations in astrophysics.
However, just at a time when, using these refined techniques,
Lorentz symmetry is being verified experimentally at a much improved
level of accuracy,
there is growing interest in theoretical models
with only an approximate Lorentz symmetry.
Part of this research is based on the realization
that certain types of phenomenologically viable
modifications of present particle-physics
models can be based on renormalizable field theories in which
indeed there is only an
approximate Lorentz symmetry~\cite{CollodayKostelecki97,ColemanGlashow}.
Perhaps even more significant is the fact that
in the quantum-gravity literature models based on an
approximate Lorentz symmetry have been recently considered
in most research lines, including models based on ``spacetime foam"
pictures~\cite{Amelino-Cameliaetal98,garayPRL}, ``loop quantum gravity"
models~\cite{GambiniPullin99,AlfraoMoralesTecotlUrrutia01},
certain ``string theory" scenarios~\cite{susskind},
and ``noncommutative geometry"~\cite{dsr1,jurekNEWdsrkmink}.

While the experimental accuracies at which Lorentz symmetry has
been so far verified are remarkable in absolute terms,
and constrain very significantly general parametrizations
of possible descriptions of Lorentz symmetry as an approximate
symmetry~\cite{CollodayKostelecki97,ColemanGlashow}, the studies motivated
by quantum gravity predict departures from Lorentz symmetry
that are naturally governed by the minute quantum-gravity length scale $L_{\rm QG}$,
which is usually identified, up to a few orders of magnitude,
with the Planck length $L_{\rm P} \sim 10^{-33}\;\hbox{cm}$
(the inverse of the huge Planck energy scale $E_{\rm P} \sim 1/L_{\rm P} \sim 10^{28} \;\hbox{eV}$) and this leads to effects that are too small for testing with present
Lorentz-symmetry laboratory tests.

Interest in tests of Planck-scale modifications of Lorentz symmetry
has also increased recently with the
realization~\cite{kifu,gacTP}
that these modifications of Lorentz symmetry
provide one of the possible solutions of the so-called ``cosmic-ray
paradox".
The spectrum of observed cosmic rays was expected to be affected
by a cutoff at the scale $E_{\rm GZK} \sim 5 {\cdot} 10^{19} \; \hbox{eV}$.
Cosmic rays emitted with energy higher than $E_{\rm GZK}$
should interact with photons in the cosmic microwave
background and loose energy by pion emission, so that
their energy should have been reduced to the $E_{\rm GZK}$ level
by the time they reach our Earth observatories.
However, the AGASA observatory
has reported several observations of cosmic rays with energies exceeding
the $E_{\rm GZK}$ limit by nearly an order of magnitude~\cite{gzkdata}.
This experimental puzzle will only be established when confirmed
by other observatories, and solutions which do not rely on Planck-scale
physics have been discussed in the literature, but
it is noteworthy that
the type of Planck-scale modification of Lorentz symmetry
described in Ref.~\cite{Amelino-Cameliaetal98}
can produce~\cite{kifu,gacTP}
an increase in the threshold energy for pion production
in collisions between cosmic rays and microwave photons,
and the increase is sufficient to explain away the puzzle
raised by the mentioned ultra-high-energy
cosmic-ray observations.

In discussions of this possible relevance of Planck-scale
(quantum-gravity) modifications of Lorentz symmetry
for the cosmic-ray paradox it is commonly assumed
that those same  modifications of Lorentz symmetry
could not be tested in controlled experiments\footnote{There is
however interest in testing them through their implications
for other classes of astrophysics
observations~\cite{domokos,Amelino-Cameliaetal98,Schaefer99,Billeretal99,glast}.}.
Indeed this is the case for all presently explored techniques
for laboratory tests of Lorentz symmetry, in which the
relevant Planck-scale effects would fall below sensitivity.
In this paper we consider the possibility
of Planck-scale Lorentz-symmetry tests in laser interferometry.
We give a schematic description of possible setups for these interferometers
which could be used to search for the relevant effects.

We take as our starting reference point LIGO/VIRGO-type~\cite{LIGO,VIRGO}
and LISA-type~\cite{LISA}
intereferometers, but we also point out that
several technological improvements should be achieved in order to
reach the required sensitivity levels in
the setups we consider.
Since (as we shall show)
the type of signal that represents a signature of the
Planck-scale effects here considered can be ``on" for a time of choice
of the experimenter (it is not a short-duration signal),
we will, at least in first instance,
focus on the level of sensitivity that advanced interferometers
can achieve in the study of a
stably-periodic signal observed
over a reference time of one year.
The ultimate sensitivity goal
of the LIGO and VIRGO interferometers
is such that short-duration (bursting) effects
inducing strain\footnote{The strain is here defined,
as conventional, according to $h \equiv \Delta L/L$,
where $L$ is the reference
length of the interferometer arms and $\Delta L$ is the change of
the length of the arms due to the effect under study.}, $h$,
at the level $h \sim 10^{-22}$ could be detected.
If the effect under study effectively induces
steady periodic variations of
the optical length of the arms of the interferometer
with period $(100\;\hbox{Hz})^{-1}$ (the strain sensitivity of LIGO/VIRGO
interferometers is at its maximum around  $100\;\hbox{Hz}$),
and if these periodic variations can be observed
for a full year, one then obtains
a year-integrated sensitivity
of order $h \sim 10^{-27}$.
[For smaller interferometers,
such as GEO600, this year-integrated sensitivity
would be~\cite{GEO600homepage} at the level $h \sim 10^{-26}$.]

In the next Section we present a naive ``back-of-the-enevelope" analysis
comparing the sensitivity of advanced interferometers
to the magnitude of the relevant Planck-scale departures
from Lorentz symmetry. Although this estimate is admittedly
simplistic, it is noteworthy that, in spite of the smallness
of the Planck length, the comparison is encouraging for the
goal of Planck-scale sensitivity.
In Section~3 we discuss  some interferometric
setups in which the Planck-scale effects
lead to a characteristic signature.
Since the relevant quantum-gravity-motivated models predict an
energy-dependent modification of the laws of light propogation,
we find appropriate to analyze our proposal also from the perspective
of a search for a photon mass \cite{GoldhaberNietoxx,LaemmerzahlHaugan01},
which would lead to the same qualitative effect\footnote{We shall not
comment instead on
models based on other types of modifications
of the Maxwell equations~\cite{CollodayKostelecki97},
in which the departures from Lorentz symmetry
are not energy dependent.}
(but, as emphasized in the following,
significant quantitative differences).
While in Sections~3 our discussion relies on an idealization
in which certain experimental limitations are ignored,
these experimental limitations are considered in Section~4,
and represent an (ambitious) agenda for those who might
be interested in pursuing the proposal we are here putting forward.
Closing remarks are offered in Section~5.

\section{Planck-scale modifications of the laws of light propagation
and advanced interferometers}

The mentioned debate on the possibility that the puzzling
observations of UHE cosmic rays might be due to a
Planck-scale deformed dispersion relation is focusing
on the first
phenomenological proposal of a quantum-gravity-motivated
dispersion relation put forward in Ref.~\cite{Amelino-Cameliaetal98}.
There it was argued that such deformed dispersion relations might
characterize quite a few approaches to the quantum-gravity problem.
In proposing a phenomenological program looking for these
new effects
it appeared natural~\cite{Amelino-Cameliaetal98,polonpap}
to start these investigations with the initial
simple {\it ansatz}\footnote{We adopt conventions
with $\hbar=c=1$.}
\begin{equation}
m^2 = E^2 -  \vec{p}^2 + f(E,\vec{p}^2;E_{\rm P})
\simeq E^2 - \vec{p}^2
+ \eta \left(\frac{E^3}{E_{\rm P}}\right)
~,
\label{displead}
\end{equation}
where $\eta$ is a dimensionless coefficient, which is usually
assumed~\cite{Amelino-Cameliaetal98}
to be $10^{-3} \le \eta \le 1$, reflecting the intuition that the
quantum-gravity scale should be somewhere between the grand unification
scale ($\sim 10^25 eV$) and the Planck scale $E_{\rm P}$.
The approximation $f(E,\vec{p}^2;E_{\rm P}) \simeq \eta E^3/E_{\rm P}$
is to be valid for $m \ll E \ll E_{\rm P}$
and assumes that the leading-order difference
between the exact dispersion relation and the ordinary classical-spacetime
dispersion relation comes in with $E/E_{\rm P}$ overall suppression factor.

Dispersion relations that are closely related to (\ref{displead})
as well as some alternative Planck-scale-deformed
dispersion relations are being discussed in
the context of the ``loop quantum gravity" theory,
in studies~\cite{GambiniPullin99,AlfraoMoralesTecotlUrrutia01}
which are motivated by the proposals
put forward in Refs.~\cite{Amelino-Cameliaetal98,kifu,gacTP}.
More recently quantum-gravity-deformed dispersion
relations were also discussed in the string-theory literature
(see, {\it e.g.}, Ref.~\cite{susskind}).
We will however focus
on the {\it ansatz} (\ref{displead}), since it is the one
being considered for solutions of the cosmic-ray
puzzle, and anyway experimental strategies which are found
to be well suited for testing (\ref{displead}) should then
be easily adaptable for testing other possible quantum-gravity-deformed
dispersion relations.

In our analysis of light beams in interferometers
we will study the relation (\ref{displead})
through its implications for the relation between
frequency, $\omega$, and wave vector, $k$.
We adopt the notation $\omega_{\rm QG}$
for the frequency scale that corresponds to the energy scale $E_{\rm P}/\eta$,
and therefore $ 10^{41} \;\hbox{Hz} \le |\omega_{\rm QG}| \le 10^{44}\;\hbox{Hz}$.
Since one could in principle consider~\cite{polonpap} both positive and
negative $\eta$,
we shall remain open to both possibilities:
our (dimensionful) parameter $\omega_{\rm QG}$ can be
positive or negative.
With this notation, from (\ref{displead}) one obtains
\begin{equation}
k \simeq \omega + \frac{1}{2}
\frac{\omega^2}{\omega_{\rm QG}} \label{Dispersion2}
\end{equation}
where $k \equiv \sqrt{{\mbox{\boldmath$k$}}^2}$.
Accordingly the group velocity for a wave packet of photons
takes the form $v_g = d\omega/dk \simeq 1 + \omega / \omega_{\rm QG}$ and
the phase
velocity $v_p = \omega/k \simeq 1 - \frac{1}{2} \omega / \omega_{\rm QG}$.

The physical mechanism that is basically exploited in the
interferometric tests here proposed originates from the fact that,
with the Planck-scale deformation of the dispersion relation,
in an interferometric setup  photons of different
energies would be affected differently by the deformation:
the Planck-scale deformation
introduces a (minute but nonnegligible) dependence of the phase
velocity on the energy/frequency of the photon.
It has already been suggested~\cite{polonpap,mynapap}
that the remarkable sensitivities of these advanced interferometers
might render them useful for studies of Planck-scale effects,
but the implications of the Planck scale for Lorentz symmetry
were not previously considered in this respect.
Postponing to the next Sections a more detailed analysis,
in this Section we just want to make a crude estimate of
the contribution to the phase of a light wave that would
come from this Planck-scale effect, and we want to compare it
with the phase-sensitivity levels that are expected to be reached
with LIGO/VIRGO-type and LISA-type interferometers.

Since the velocity\footnote{We are here loosely referrring to
the speed of the wave, since an effect of order $\omega / \omega_{\rm QG}$
affects both the phase velocity and the group velocity in the
framework we are here considering.
In the more detailed analysis reported in the Section~3 we express
everything directly in terms of $k$ and $\omega$,
rather then velocities.}
is modified at the level $\omega / \omega_{\rm QG}$
one can easily estimate (and this is confirmed by our analysis
in the next Sections) that the corresponding Planck-scale contribution
to the phase would be at the
same level $\phi_{\rm QG} =\phi_{\rm noQG} (1+\omega / \omega_{\rm QG})$,
and this would effectively correspond to a change in our
estimate of the optical length $L$ of the interferometer
which is again at the same $\omega / \omega_{\rm QG}$
level: $L_{\rm QG} =L (1+\omega / \omega_{\rm QG})$.
Since for visible
light $10^{-28} \le |\omega / \omega_{\rm QG}| \le 10^{-25}$
one estimates $10^{-28} \le |L_{\rm QG}-L|/L \le 10^{-25}$.

We want to compare this crude characterization of the magnitude
of the Planck-scale effects with the
expected sensitivity of LIGO/VIRGO- and LISA-type intereferometers.
The most publicized sensitivity characterization for these
interferometers is the one for bursting/short-duration
gravity-wave signals.
For an incoming gravity wave of 100 Hz frequency (ideal for the
LIGO/VIRGO setup, with optical length $\sim 1000 \;\hbox{km}$)
the amplitude/strain of a short-duration gravity wave
must be at least at the level $h \sim 10^{-22}$
in order for it to be revealed by the advanced phase~\cite{ligoadvanced}
of LIGO/VIRGO interferometers.
As mentioned in the Introduction, if instead of a bursting signal
the interferometer is affected by a steady gravity wave of period 100 Hz,
which remains steady for a full year, one should consider the year-integrated
strain sensitivity, which would be at the level $h \sim 10^{-27}$.
Since $h$ can be seen as $|L_{\rm GW}-L|/L$, where $L$ is the optical length of
the arms of the interferometer when not affected by a gravity wave
and $L_{\rm GW}$ is the maximum (or minimum) optical length of
the arms of the interferometer in presence of the gravity wave,
this sensitivity at the $h \sim 10^{-27}$ level can be encouraging
for the search of quantum-gravity effects at the
level $10^{-28} \le |L_{\rm QG}-L|/L \le 10^{-25}$.

For LISA similar or better sensitivity to the effects we are here considering
can be expected.
In searches of high-frequency ({\it e.g.} $100 \;\hbox{Hz}$)
gravity waves, LISA's remarkable phase sensitivity does not
translate into an equally remarkable strain sensitivity
because the key advantage of a LISA-type interferometer,
its huge arms length ($L \sim 10^{10}\;\hbox{m}$),
is not fully exploited in presence of high-frequency
gravity waves. In fact, for high-frequency gravity waves the linearity between
arms length and observed phase difference is lost.
Instead in our searches of the effects
induced by a Planck-scale deformation of
the dispersion relation the relevant phase differences are always
proportional to the arms length, and LISA's remarkable phase sensitivity
can be fully exploited.

Of course, the estimates we are providing in this section are crude
and idealized. It is somewhat encouraging that at least at this level
the suppression induced by the smallness of the Planck length
does not appear to be unsurmountable, since, even at the same crude level
of analysis, most other experimental setups would immediately prove
to be inadequate for Planck-scale studies).
In the next Section we discuss some interferometric setups that could
be used to find evidence of the Planck-scale effects we are considering.
This will provide the basis for our more realistic discussion, in Section~4,
of the challenges that must be faced in order to render meaningful
the encouraging naive estimate we obtained here.

As a corollary, in our analysis we also consider possible applications
of our strategy in searches of a photon mass.
Since a photon mass would affect the dispersion relation for light waves,
the effect is qualitatively analogous to the one we are
considering from the Planck-scale perspective.
However, at the quantitative level the two effects are clearly
distinguishable.
In presence of a nonvanishing photon mass the standard $k = \omega$
dispersion relation takes
the form\footnote{And correspondingly group and phase velocity take
the form $v_g \approx 1- m^2/(2 \omega^2)$
and $v_p = 1 + m^2/(2 \omega^2)$
respectively.}
\begin{equation}
k \approx \sqrt{\omega^2 - m^2}
\approx \omega - \frac{m^2}{2 \omega} ~.
\end{equation}
When the Planck-scale effect is present (and there is no photon mass)
the deformation of the dispersion relation becomes more and more
significant as the frequency of the wave increases.
The opposite is true for the case of a photon mass (and no Planck-scale
effects):
the photon-mass deformation of the dispersion relation becomes more and more
significant as the frequency of the wave decreases.

\section{Proposal of interferometric setups for a Planck-scale-induced
phase difference}

In the preceding Section we just compared
the ``phase sensitivity" of advanced interferometers
to the magnitude of the contribution to the phase of a light wave
due to the Planck-scale effects we are investigating.
In light of the tempting result of this comparison we are encouraged
to look for ways to
test Planck-scale-deformed dispersion relations using
laser-light interferometers.

There are probably a large number of interferometric setups that
could be considered for our quantum-gravity tests.
To render our proposal more
specific we will discuss in some detail two such interferometric
setups.
A common feature of our interferometric setups
is that they involve two beams at different energies/frequencies.
As we shall show,
the energy-dependence characteristic of Planck-scale
deformed dispersion relations is such that the presence
of beams at different energies in the interferometer
naturally gives rise to Planck-scale-dependent phase differences.
We specifically consider two frequencies, a reference frequency $\omega$
and the doubled frequency $2 \omega$, but interferometric setups
that exploit other types of pairs of frequencies should be possible.
We choose to consider frequency doubling because of the relatively
wide availability of second harmonic generation (SHG)
or ``frequency doublers'' (see e.g. \cite{Sauter96}).

The discussion in this Section
is ``idealized'', largely ignoring various potential challenges
that would be encountered
in actual realizations of our interferometric setups.
As announced, we will comment on some of these challenges in Section~4.

\subsection{Phase difference through splitting in
energy and configuration space}

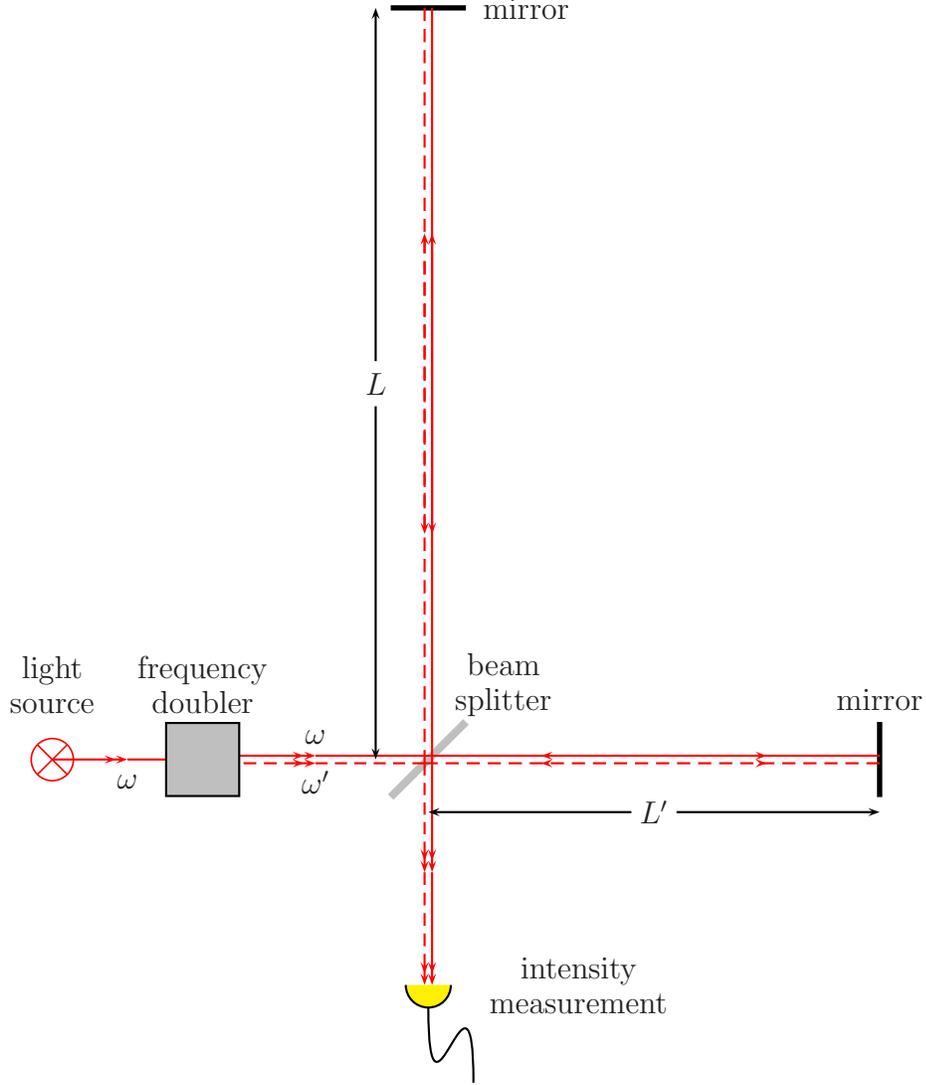
\begin{figure}[t!]
\begin{center}
\begin{pspicture}(-6,-5)(6,10)
\psline[linewidth=3pt,linecolor=lightgray](-0.5,-0.5)(0.5,0.5)
\psline[linecolor=red]{->>}(-5,0)(-4,0)
\psline[linecolor=red]{->>}(-4,0)(-3,0)
\psline[linecolor=red]{->>}(-3,0.05)(-1.5,0.05)
\psline[linecolor=red,linestyle=dashed]{->>}(-3,-0.05)(-1.5,-0.05)
\psline[linecolor=red](-1.5,0.05)(0.05,0.05)
\psline[linecolor=red,linestyle=dashed](-1.5,-0.05)(-0.05,-0.05)
\psframe[fillstyle=solid,fillcolor=lightgray](-3.5,-0.5)(-2.5,0.5)
\rput(-5,0){\color{red} \huge $\otimes$}
\rput(-5,1){\txt{light \\ source}}
\rput(-3,1){\txt{frequency \\ doubler}}
\rput(-4,-0.3){$\omega$}
\rput(-1.5,-0.3){$\omega^\prime$}
\rput(-1.5,0.3){$\omega$}
\rput(1,1){\txt{beam \\ splitter}}
\rput(1.3,10){mirror}
\rput(6,0.8){mirror}
\psline[linewidth=2pt](-0.5,10)(0.5,10)
\psline[linewidth=2pt](6,-0.5)(6,0.5)
\psline[linecolor=red]{->}(0.05,0.05)(0.05,7)
\psline[linecolor=red,linestyle=dashed]{->}(-0.05,-0.05)(-0.05,7)
\psline[linecolor=red]{->}(0.05,10)(0.05,3)
\psline[linecolor=red,linestyle=dashed]{->}(-0.05,10)(-0.05,3)
\psline[linecolor=red]{-}(0.05,0.05)(6,0.05)
\psline[linecolor=red,linestyle=dashed]{-}(-0.05,-0.05)(6,-0.05)
\psline[linecolor=red]{->}(4.4,0.05)(4.5,0.05)
\psline[linecolor=red,linestyle=dashed]{->}(4.4,-0.05)(4.5,-0.05)
\psline[linecolor=red]{->}(1.6,0.05)(1.5,0.05)
\psline[linecolor=red,linestyle=dashed]{->}(1.6,-0.05)(1.5,-0.05)
\psline[linecolor=red]{->>}(0.05,0.05)(0.05,-1.5)
\psline[linecolor=red,linestyle=dashed]{->>}(-0.05,-0.05)(-0.05,-1.5)
\psline[linecolor=red]{->>}(0.05,-1.5)(0.05,-3)
\psline[linecolor=red,linestyle=dashed]{->>}(-0.05,-1.5)(-0.05,-3)
\psarc[fillstyle=solid,fillcolor=yellow](0,-3){0.3}{180}{360}
\psbezier(0,-3.3)(0,-5.3)(0.6,-2.3)(0.6,-4.3)
\rput(2,-3){\txt{intensity \\ measurement}}
\psline{->}(2.7,-0.7)(0,-0.7)
\psline{->}(3.3,-0.7)(6,-0.7)
\rput(3,-0.7){$L^\prime$}
\psline{->}(-0.7,5.3)(-0.7,10)
\psline{->}(-0.7,4.7)(-0.7,0)
\rput(-0.7,5){$L$}
\end{pspicture}
\end{center}
\caption{The unequal--arm interferometer
which is capable to search for dispersion effects.
The paths for the original ($\omega$)
and the frequency doubled ($\omega^\prime$) beams
photons are identical in configuration space, but we draw separate
lines for conceptual clarity.
\label{GWInterferometer}}
\end{figure}

Let us consider an interferometer with LIGO/VIRGO or LISA-type
setup (and dimensions) with two orthogonal arms,
respectively of length $L$ and $L^\prime$ (see Fig.~\ref{GWInterferometer}).
We keep $L$ and  $L^\prime$ distinct because our signal will turn out
to be proportional to $|L - L^\prime|$.
This will come at some cost in the sensitivity of the
interferometer, but probably not more\footnote{See, {\it e.g.},
Ref.~\cite{LarsonHellingsHiscock02} for a comparison
between an unequal--arm interferometer and a comparable
equal--arm interferometer. Indeed,
the analysis reported in Ref.~\cite{LarsonHellingsHiscock02}
leads to the conclusion that the unequal--arm interferometer
has sensitivity reduced by about a factor 10 with
respect to the equal--arm interferometer.} than a factor 10.

The scheme of this interferometric setup is shown
in  Fig.~\ref{GWInterferometer}.
Before entering the interferometer, a monochromatic wave with
frequency $\omega$
goes through a frequency doubler.
Both emerging beams, of frequencies $\omega$ and $2 \omega$,
are then split into a part that goes through the arm
of length $L$ and a part that goes through the arm of length $L^\prime$.
When the beams are finally back (after reflection by a mirror)
at the point where the interference patterns are formed,
one then has access to two interference patterns: one
interference pattern combines two waves of frequency $\omega$
and the other intereference pattern is formed combining
analogously two waves of frequencies $2 \omega$.
In an idealized setup (ignoring for example a possible wavelength
dependence in beam-mirror interactions)
the observed intensities\footnote{In practice,
it may be convenient to use frequency filters in such a way
that at any given instant only one of the two intensities
is observed. It might also be possible to insert another
beam splitter into the beam going to the apparatus measuring
the intensity. Both resulting beams may be frequency analyzed,
one for the frequency $\omega$, the other for $2\omega$.
By measuring the intensity of both of these beams
both intensities may be measured simultaneously.}
would be governed by
\begin{equation}
I_{\omega} \propto
\frac{1}{2} \left(1
+ \cos\phi_{\omega} \right) \, ,
\quad \phi_{\omega} = k (L^\prime - L) \, ,
\label{phase1}
\end{equation}
\begin{equation}
I_{2 \omega} \propto
\frac{1}{2} \left(1
+ \cos\phi_{2 \omega}\right) \, ,
\quad \phi_{2 \omega} = k^\prime (L^\prime - L) \, ,
\label{phase2}
\end{equation}
where $k^\prime$ is the wavelenght associated
with the doubled frequency $2 \omega$.
In Minkowski spacetime,
with its standard dispersion relation, one has that $k^\prime = 2 k$,
but in the case of the Planck-scale-deformed dispersion relation
\begin{equation}
k^\prime \simeq 2 k + \frac{k^2}{\omega_{\rm QG}}
\, .
\label{kkprime}
\end{equation}

We can then rewrite $\phi_{2 \omega} - \phi_{\omega}$, from (\ref{phase1}) and
(\ref{phase2}), using again  the Planck-scale-deformed dispersion relation
\begin{equation}
\phi_{2 \omega} - \phi_{\omega}
= k^\prime (L^\prime - L) - k (L^\prime - L)
= \omega
\left(1 + \frac{3}{2} \frac{\omega}{\omega_{\rm QG}}\right) (L^\prime
- L) \, .
\label{phasediffGAC}
\end{equation}

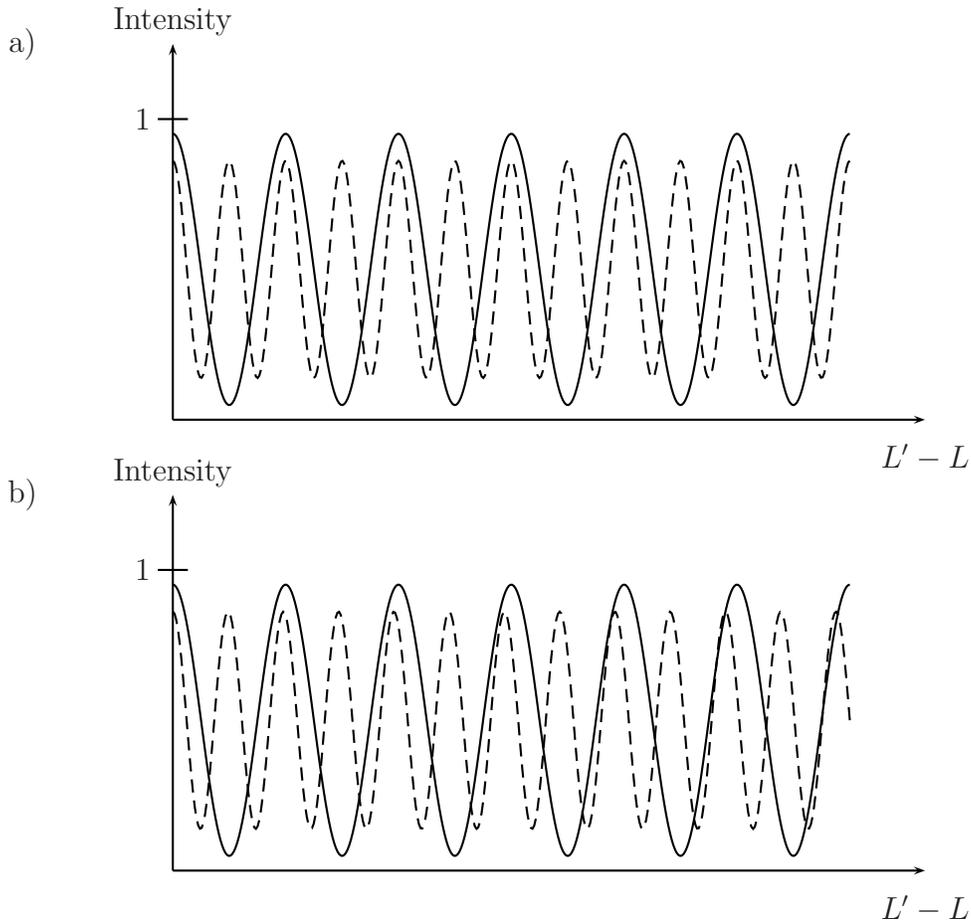
\begin{figure}[t!]
\begin{center}
\begin{pspicture}(-1,0)(10,11)
\put(0,6){
\rput(-2,5){a)}
\psline{->}(0,0)(0,5) 
\rput(0,5.3){Intensity}
\psline(-0.2,4)(0.2,4)
\rput(-0.4,4){1}
\rput(10,-0.5){$L^\prime - L$}
\psline{->}(0,0)(10,0) 
\psset{plotpoints=400}
\psplot{0}{9}{x 120 mul cos 1.9 mul dup mul 0.195 add}
\psplot[linestyle=dashed]{0}{9}{x 240 mul cos 1.7 mul dup mul 0.555 add}
}
\rput(-2,5){b)}
\psline{->}(0,0)(0,5) 
\rput(0,5.3){Intensity}
\psline(-0.2,4)(0.2,4)
\rput(-0.4,4){1}
\rput(10,-0.5){$L^\prime - L$}
\psline{->}(0,0)(10,0) 
\psset{plotpoints=400}
\psplot{0}{9}{x 120 mul cos 1.9 mul dup mul 0.195 add}
\psplot[linestyle=dashed]{0}{9}{x 245 mul cos 1.7 mul dup mul 0.555 add}
\end{pspicture}
\end{center}
\caption{{\bf a)} Qualitative description
of the dependence
on $L'-L$ of the intensities $I_{\omega}$
and $I_{2 \omega}$, according to ordinary Lorentz symmetry.
We show different maximum values for $I_{\omega}$
and $I_{2 \omega}$ to reflect the fact that the intensities
of the two beams that emerge from SHG in general are not identical.
However, this is not of concern for us: independently of
the values of the maximum intensities
exact Lorentz symmetries introduces a correlation
between the values of $I_{\omega}$
and $I_{2 \omega}$ at given values of $L^\prime - L$.
For example, whenever $L'-L$ is such that there is a
maximum of  $I_{2 \omega}$ there  must also necessarilly be
a maximum or a minimum of $I_{\omega}$.
{\bf b)} Qualitative description
of the dependence
on $L'-L$ of the intensities $I_{2 \omega}$
and $I_{\omega}$, according to the quantum-gravity-induced
departure from Lorentz symmetry here considered.
By comparing with a) the implications of the Planck-scale
effects are visible. For example, one notices that the quantum-gravity
effect induces a misalignment between the maxima of  $I_{2 \omega}$
and the maxima and minima of $I_{\omega}$. As shown in our
analysis this misalignment is proportional
to $(\omega^2/\omega_{\rm QG}) (L^\prime - L)$.} \label{weaveOUTofPHASE}
\end{figure}

A proper description of
the meaning of this phase-difference relation requires a few considerations.
This phase-difference relation characterizes a key difference
between the standard dispersion relation and the
Planck-scale-deformed dispersion relation in the interferometric
setup we are considering.
In the case of the standard classical-spacetime dispersion relation
one expects, as illustrated in Fig.~\ref{weaveOUTofPHASE}a,
a specific type of correlations,
which follow straightforwardly from $k^\prime =2 k$,
between the values of $I_{\omega}$
and $I_{2 \omega}$ for given values of $L^\prime - L$.
For example, clearly one expects that
the intensity $I_{2 \omega}$ of
the wave at frequency $2 \omega$ has a maximum
whenever $L^\prime - L = 2 n \pi/\omega$
(with $n$ any integer number),
and that correspondingly the intensity $I_{\omega}$ of
the wave at frequency $\omega$ has either a maximum
or a minimum.
One therefore predicts,
without any need to establish
the value of $n$, that the
configurations in which there is a maximum of $I_{2 \omega}$
must also be
configurations in which there is
a maximum or a minimum of $I_{\omega}$, see Fig.~\ref{weaveOUTofPHASE}a.
The Planck-scale-deformed dispersion relation modifies this
prediction: for example, as codified by Eq.~(\ref{phasediffGAC}),
the dispersion relation (\ref{Dispersion2})
predicts that configurations in which there is
a maximum of $I_{2 \omega}$
should be such that $I_{\omega}$
is in the neighborhood but not exactly at
one of its maximum/minimum values (see Fig.~\ref{weaveOUTofPHASE}b).
More precisely, when $L^\prime - L$ is such
that $I_{2 \omega}(L^\prime - L)$ is at a maximum value
the quantum-gravity effect predicts
that $I_{\omega}(L^\prime - L)$ should differ from
a maximum/minimum value of $I_{\omega}$ as if
for being out-of-phase
by an amount
\begin{equation}
\phi_{2 \omega} - \phi_{\omega}
= \frac{3}{2} \frac{\omega^2}{\omega_{\rm QG}} (L^\prime
- L) \, .
\label{phasediffnew}
\end{equation}

The comparison of maxima/minima of the functions $I_{2 \omega}(L^\prime - L)$
and $I_{\omega}(L^\prime - L)$ is not necessarily the best\footnote{This
point was emphasized to us by A.~R\"{u}diger.} feature
of the general comparison between  $I_{2 \omega}(L^\prime - L)$
and $I_{\omega}(L^\prime - L)$ from the point of view of experimental
searches, because close to a stationary point of course functions
vary very slowly (only quadratic term is effective).
But the standard classical-spacetime picture established a connection
between the  $I_{2 \omega}(L^\prime - L)$
and $I_{\omega}(L^\prime - L)$  graphs even away from the stationary points.
Moreover one could easily revise this setup
in such a way to have $\omega$ and $3 \omega$ beams
(rather than $\omega$ and $2 \omega$ beams)
in which case standard classical-spacetime picture established
that some points of maximum positive slope of $I_{3 \omega}(L^\prime - L)$
should correspond to points of maximum (minimum) slope
of $I_{\omega}(L^\prime - L)$.

This type of characteristic feature would be easily looked for by, for example,
taking data at values of $L'-L$ that differ from one another
by small (smaller than $1/\omega$)
amounts in the neighborhood of a value of $L'-L$ that corresponds, say,
to a maximum of $I_{2 \omega}$.
Perhaps, techniques for the active control of mirrors which are already
being used in modern intereferometers might be adapted
for this task, and the development of dedicated techniques
does not appear beyond our reach.

In closing this Subsection let us return to Eqs.~(\ref{phase1})
and (\ref{phase2}) and remove one of the key idealizations
in those formulas. As mentioned, it appears likely that
in addition to the contributions $k (L^\prime - L)$
and $k^\prime (L^\prime - L)$,
respectively to $\phi_{\omega}$ and $\phi_{2 \omega}$,
there would also be some different $(L^\prime - L)$-independent
contributions to the phases due for example to the different
response of the mirrors (and the beam splitter)
to the $\omega$ wave and the $2 \omega$ wave.
Such a phase difference would itself induce a
misalignment between the
maxima of  $I_{2 \omega}(L^\prime - L)$
and the maxima and minima of $I_{\omega}(L^\prime - L)$,
which could of course have negative implications for the
observability of the analogous quantum-gravity effect.
One might try to determine these unwanted phase
differences  to high precision in the laboratory before
carrying out the final experiment described above. In alternative
one could consider the possibility of exploiting
the fact that the misalignment due to
the quantum-gravity effect is proportional to $(L^\prime - L)$
while the other potential source of misalignment
is $(L^\prime - L)$-independent. One could for example repeat the
entire procedure necessary to establish the amount of misalignment
between the maxima of  $I_{2 \omega}(L^\prime - L)$
and the maxima and minima of $I_{\omega}(L^\prime - L)$
for two macroscopically different values of $(L^\prime - L)$.
If for the two macroscopically different values of $(L^\prime - L)$
one found the same amount of misalignment the quantum-gravity
effect would be excluded, whereas in the opposite case one would
have evidence in favour of the quantum-gravity effect.

\subsection{Interferometry in energy space}

As another possible realization of an interferometric setup that
is sensitive to deformations of
the dispersion relation we propose a setup in which the
frequency (or energy) is the parameter characterizing the splitting
of the photon state. The main experimental procedure for doing
that is SHG. (Again one may consider a more general setup where
the frequency may have other than doubled values.)
If an incoming wave has a frequency $\omega$ then,
after passing through the frequency doubler, the outgoing wave
in general consists of two components, one possessing
a frequency $\omega$ and the other a frequency $2 \omega$.

\begin{figure}[t!]
\begin{center}
\begin{pspicture}(0,-1)(10,4)
\psline{<->}(0,4)(0,0)(10,0)
\rput{90}(-0.3,3.5){frequency}
\rput(9.5,-0.3){position}
\psline[linecolor=red]{->}(0,1)(3,1)
\put(0.2,1.2){$u_0(x,t) \sim e^{- i \omega t}$}
\psframe[fillstyle=solid,fillcolor=lightgray](3,0)(3.5,3.8)
\rput(3.25,-0.3){$x = 0$} \rput(3.25,4.3){SHG}
\psline[linecolor=red]{->}(3,1)(3.5,1)
\psline[linecolor=red]{->}(3,1)(3.5,2)
\psline[linecolor=red]{->}(3.5,1)(6,1)
\psline[linecolor=red]{->}(3.5,2)(6,2)
\put(4.2,1.2){$\sim e^{- i \omega t} $}
\put(4.2,2.2){$\sim e^{- 2 i \omega t} $}
\psframe[fillstyle=solid,fillcolor=lightgray](6,0)(6.5,3.8)
\rput(6.25,-0.3){$x = L$} \rput(6.25,4.3){SHG}
\psline[linecolor=red,linestyle=dashed]{->}(6,1)(6.5,1)
\psline[linecolor=red]{->}(6,1)(6.5,2)
\psline[linecolor=red]{->}(6,2)(6.5,2)
\psline[linecolor=red,linestyle=dotted]{->}(6,2)(6.5,3)
\psline[linecolor=red,linestyle=dotted]{->}(6,1)(6.5,3)
\psline[linecolor=red,linestyle=dashed]{->}(6,2)(6.5,1)
\psline[linecolor=red,linestyle=dashed]{->}(6.5,1)(9,1)
\psline[linecolor=red]{->}(6.5,2)(10,2)
\psline[linecolor=red,linestyle=dotted]{->}(6.5,3)(9,3)
\put(7.2,1.2){$\sim e^{- i \omega t}$}
\put(7.2,2.2){$u_2^{(2\omega)}(x,t) \sim e^{- 2 i \omega t}$}
\put(7.2,3.2){$\sim e^{- 3 i \omega t}$ (forbidden)}
\end{pspicture}
\end{center}
\caption{The waves leaving two frequency doublers.
The two waves leaving the second doubler with
frequency $2\omega$ form an interferometer since
they move between the two doublers at different frequencies.
Waves not contributing to our interferometer are displayed
by dashed lines. Dotted waves are forbidden due to the
matching condition. Note that the different paths in the figure
only distinguish different frequencies; all beams follow the same
path in configuration space.
 \label{SHGInterferometer}}
\end{figure}
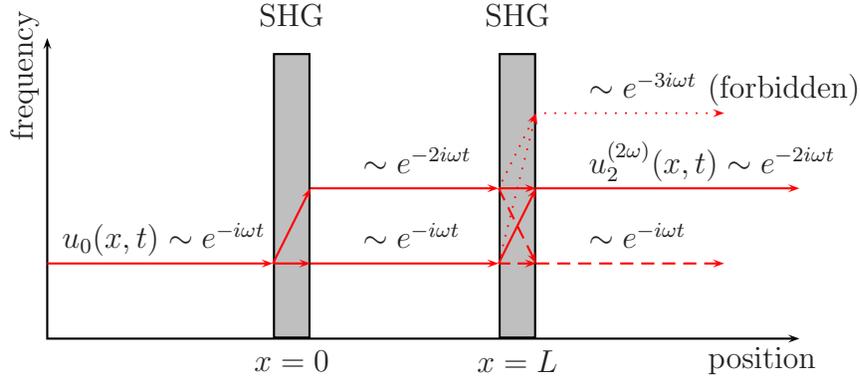

We intend to consider an interferometric setup, described
in Fig.~\ref{SHGInterferometer},
where two SHG
are positioned in a row at a distance $L$. After the second SHG
the photonic state consists of two components with
frequency $\omega$ and two components with frequency $2 \omega$.
The first contribution to the emerging $2 \omega$ beam is the
transmission of the component which left the first SHG
as a $2 \omega$ wave, while the other
contribution to the emerging $2 \omega$ beam is the
doubled component of that part which went
through the first SHG without change in the frequency.
Therefore, the final $2 \omega$ component represents an
interferometer in energy space.
Similarly there is contribution
to the emerging $\omega$ beam which is the
transmission of the component which left the first SHG
still as a $\omega$ wave (part of the beam which is not affected
by any of the two SHG), while the other
contribution to the emerging $\omega$ beam is the
halved component of a part of the beam
which left the first SHG as a $2 \omega$ wave.
Therefore, also the final $\omega$ component represents an
interferometer in energy space.
(All beams travel always along the same path in configuration
space. There is no need for a beam splitting in configuration
space. Beam splitting takes place with respect to energy.)

We use a general model in order to describe the process of
frequency doubling, or more generally, addition and subtraction
of frequencies. This is called second harmonic generation, SHG.
Thereby, light enters a nonlinear material, that is, a material
whose dielectric function depends on the frequency. Thus, inside
the material, the polarization depends on the square of the
incoming electric field. Consequently, if the incoming wave
consists of two monochromatic parts
\begin{equation}
u_0(x, t) = A_{01} e^{i \alpha_{01}} e^{i (k(\omega_1) (x - x_0)
- \omega_1 t)} + A_{02} e^{i \alpha_{02}} e^{i (k(\omega_2)
(x - x_0) - \omega_2 t)}
\end{equation}
then in the nonlinear medium the polarization field consists
of components possessing the frequencies $\omega_{\rm shg}
= \{2 \omega_1, 2 \omega_2, \omega_- = \omega_1 - \omega_2, \omega_+
= \omega_1 + \omega_2\}$,
and the corresponding wave
vectors $k_{\rm shg} = \{2 k(\omega_1), 2 k(\omega_2), k(\omega_1)
+ k(\omega_2), k(\omega_1) - k(\omega_2)\}$, see e.g. \cite{Sauter96}.
However, due to possible destructive interference not all of these
waves survive. (The reason for that is that the frequencies and
the corresponding wave vectors of the polarization field do not
fulfill a wave equation and thus possess a periodicity different
from the propagating waves.)  The destructive interference of
these waves is prohibited only if a phase matching
condition $k(\omega_{\rm shg}) - k_{\rm shg} = 0$
for the corresponding pair of frequencies and wave
vectors, is fulfilled. This is a condition on the
refraction index of the medium.
If this condition is not met exactly,
then this will only influence the intensities (rather than the frequencies)
of the beams leaving the SHG only.

The combinations we are interested in
are doubling, $\omega + \omega \rightarrow 2 \omega$, and
subtraction, $2 \omega - \omega \rightarrow \omega$.
It turns out, that for these two processes the phase
matching conditions are identical. The condition for
a further process $2\omega + \omega$ in general cannot
be fulfilled at the same time. The frequency doubler
only converts a part of the incoming beam to a beam
with twice the original frequency. From this it is
clear that a SHG acts as a beam splitter in frequency space and two SHGs
in a row act as an interferometer
in frequency space (Fig.~\ref{SHGInterferometer}).

Based on these considerations, we propose the following procedure:
A monochromatic plane wave
\begin{equation}
u_0(x, t) = A_0 e^{i \alpha_0} e^{i (k (x - x_0) - \omega t)}
\end{equation}
with a (real) amplitude $A_0$ and an extra phase $\alpha_0$ enters
the first SHG placed at position $x_1$.
The outcome is a frequency-doubled wave
and a part which just goes through the crystal:
\begin{eqnarray}
u_1(x, t) & = & A_{1} e^{i \alpha_{1}} e^{i \alpha_0}
e^{i k (x_1 - x_0)} e^{i (k (x - x_1) - \omega t)} \nonumber\\
& & \qquad + B_{1} e^{i \beta_{1}} e^{2 i \alpha_0}
e^{2 i k (x_1 - x_0)} e^{i (k^\prime (x - x_1) - 2 \omega t)} \label{SHG1}
\end{eqnarray}
where $k^\prime$ again is related to $2\omega$ through the modified
dispersion relation
and $A_1$, $B_1$, $\alpha_1$, and $\beta_1$ are real
functions of the amplitude and the frequency of the incoming wave.

These two waves enter the second identical SHG placed
at position $x_2 = x_1 + L$. With the phase matching
condition, only waves with frequency $\omega$
and $2 \omega$ come out. Each of these two
frequencies consists of two parts, one from
the wave which went through the material,
and the other which is either frequency
added (doubled) or frequency subtracted:
\begin{eqnarray}
u_2(x, t) & = & {\cal A} e^{i (k (x - x_2) - \omega t)} + {\cal B}
e^{i (k^\prime (x - x_2) - 2 \omega t)}\, , \label{SHG2}
\end{eqnarray}
with
\begin{eqnarray}
{\cal A} & = & A_{21} e^{i \alpha_{21}} e^{i (\alpha_{1}
+ \alpha_0 + k (x_1 - x_0))} e^{i k (x_2 - x_1)}  \nonumber\\
& &   + A_{22} e^{i \alpha_{22}} e^{i (\beta_{1}
+ 2 \alpha_0 + 2 k (x_1 - x_0) - (\beta_{11}
+ \alpha_0 + k (x_1 - x_0)))} e^{i (k^\prime (x_2 - x_1) - k (x_2 - x_1))} \\
{\cal B} & = & B_{21} e^{i \beta_{21}}
e^{i (\beta_{1} + 2 \alpha_0 + 2 k (x_1 - x_0))}
e^{i k^\prime (x_2 - x_1)} \nonumber\\
& &   + B_{22} e^{i \beta_{22}} e^{2 i (\alpha_{1}
+ \alpha_0 + k (x_1 - x_0))} e^{2 i k (x_2 - x_1)}\, . \label{SHG2bis}
\end{eqnarray}
Here all amplitudes $A_{2i}$ and $B_{2i}$ as well
as all phases $\alpha_{2i}$ and $\beta_{2i}$
depend on the amplitudes and phases of the waves entering the
second SHG only and {\it not} on the distance between the two SHGs.

By means of a frequency filter, we can select the
doubled frequency component which is
\begin{eqnarray}
u_2^{(2\omega)}(x, t) & = & \left(B_{21}
e^{i \beta_{21}} e^{i (\beta_{1} + 2 \alpha_0
+ 2 k (x_1 - x_0))} e^{i k^\prime (x_2 - x_1)} \right. \nonumber\\
& & \left.  + B_{22} e^{i \beta_{22}}
e^{2 i (\alpha_{1} + \alpha_0 + k (x_1 - x_0))}
e^{2 i k (x_2 - x_1)}\right) e^{i (k^\prime (x - x_2) - 2 \omega t)}\, .
\end{eqnarray}
The interference pattern shows up in the
intensity $I^{(2\omega)} = |u_2^{(2\omega)}(x, t)|^2$ which is
\begin{equation}
I^{(2\omega)} = B_{21}^2 + B_{22}^2 + 2 B_{21} B_{22} \cos\phi^{(2\omega)} ~,
\end{equation}
with
\begin{eqnarray}
\phi^{(2\omega)} & = & \beta_{21} + \beta_{1} + 2 \alpha_0
+ 2 k (x_1 - x_0) + k^\prime (x_2 - x_1) - \beta_{22} - 2 (\alpha_1
+ \alpha_0 + k (x_1 - x_0)) - 2 k (x_2 - x_1) \nonumber\\
& = & \beta_{21} - \beta_{22} + \beta_1 - 2 \alpha_1
+ (k^\prime - 2 k) (x_2 - x_1) \, ,
\label{good}
\end{eqnarray}
where we set $x = x_2$. Since the generation of doubled
frequencies is not very effective (may be up to 30\%), $B_{22}$ is of the order of $B_{21}$,
so that the visibility of the $2\omega$--interference
pattern may reach 1.

If we select the $\omega$--component, then we get
\begin{equation}
I^{(\omega)} = A_{21}^2 + A_{22}^2 + 2 A_{21} A_{22} \cos\phi^{(\omega)} ~,
\end{equation}
with
\begin{equation}
\phi^{(\omega)} = \alpha_{21} + \alpha_1
- \alpha_{22} - \beta_1 + \beta_{11} - (k^\prime - 2 k) (x_2 - x_1) \, .
\end{equation}
Selection of the frequency $\omega$ part is possible,
and maybe useful to improve the quality of the analysis
({\it e.g.} as an opportunity to double-check the results
obtained with the $2 \omega$-wave interference studies),
but of course
the contrast of the $\omega$-wave
interference pattern is by far not as good as for the doubled part.
(The $\omega$-wave which just
goes through the two SHG is dominating
with respect to the $\omega$-wave which is affected twice
by the SHG: $A_{21} \gg A_{22}$.)

On the basis primarily of Eq.~(\ref{good}), for the interference of
the doubled-frequency waves, we do have in this setup sentivity
to the Planck-scale deformation of the dispersion relation,
which is encoded in the
term $(k^\prime - 2 k) L$,
\begin{equation}
k^\prime - 2 k = \frac{\omega^2}{\omega_{\rm QG}} \, .
\end{equation}
However, the  term $(k^\prime - 2 k) L$ might be obscured
by the additional phases in Eq.~(\ref{good}). This can be avoided
exploiting the fact that
the other phases
in Eq.~(\ref{good}) do {\it not} depend on the distance $L$ between the SHGs.
It is therefore again useful to introduce
some form of controlled variation of $L$ in the interferometric
setup.

\subsection{Associated photon-mass analysis}

The examples of interferometric setups which we discussed in the
preceding two Subsections are in general sensitive to any type
of modification of the $\omega (k)$ dispersion relation.
In particular, in addition to the study of possible Planck-scale-induced
effects on which we focused, our interferometric setups are also
sensitive to the modification of the $\omega (k)$ dispersion relation
that would be induced by a hypothetical photon mass $m_\gamma$.

The analysis of the photon-mass
scenario can be done in complete analogy with
the one reported in the preceding two Subsections, with the
only difference that in the case of a photon mass
one finds
\begin{equation}
k^\prime - 2 k = \frac{3}{4} m_\gamma^2/\omega \, .
\label{gacphotonm}
\end{equation}
rather than the Planck-scale effect
$k^\prime - 2 k = \omega^2/\omega_{\rm QG}$
which we have been considering so far.

Clearly the two candidate effects, the ones due to Planck-scale physics
and the ones due to a photon mass, can be easily distinguished
(they would not represent significant backgrounds for one another).
If anomalous results are found in
interferometric studies of the type we are proposing
it will be easy to establish whether the anomaly is due
to the Planck-scale effects or to a photon mass.
In fact,
that Planck-scale modifications of the dispersion relation
lead to an effect which is proportional to the frequency
of the laser source, while the photon-mass-induced
effect decreases as the frequency is increased.

If interferometers working with light
frequencies of order, say, $10^{15}\;\hbox{Hz}$,
achieved, following the strategy here outlined,
sensitivity to the quantum-gravity deformed dispersion
relation, this would correspond to sensitivity to a photon
mass of order $m_{\gamma} \sim 10^{-47}\;\hbox{g}$,
3 orders of
magnitude better than recent astrophysical analyses~\cite{Schaefer99}
and one order of magnitude worse\footnote{This is actually not surprising: experiments
such as the ones discussed in Ref.~\cite{Luoetal03}
are not propagation experiments, but rather they infer limits
on the photon mass on the basis of measurements of processes
associated with the electromagnetic interactions.
Gravitational interactions are much weaker than electromagnetic
ones, and therefore a similar strategy would not be fruitful
in searches of quantum-gravity effects (which might well be seen first
in propagation studies and only at a much weaker level in
contexts involving interactions).}
than the one
obtained in laboratory experiments based on
a dynamic torsion balance~\cite{Luoetal03}.

\section{Key challenges for achieving the needed
experimental accuracy}
Interferometric techniques have improved continuously and
substantially from decade to decade from the last quarter
of the 19th century to the present times. Sensitivities that
appeared inimmaginable at one point in this development
were eventually reached and surpassed. An example that is presently
at the center of the attention of the scientific community
is the one of the interferometric searches of gravity waves:
we expect to detect gravity waves within a few years, but this
will be achieved through remarkable sensitivity improvements
with respect to the sensitivities actually achievable when the first
ideas on such studies appeared in the literature of the 1960s
and 1970s.
The proposal we are putting forward in this paper is to be intended
in the same spirit as those early discussions of interferometric
detection of gravity waves: interferometric studies of
Planck-scale deformations of the dispersion relation
may require a significant improvement of the interferometric
techniques with respect to the ones that are presently available,
but this improvement does not appear to be beyond reach,
if indeed our mastery in interferometry keeps progressing
at the pace of these past decades.

From this perspective we should stress here at least a few
of the key challenges that must be faced in attempting to
realize interferometric setups of the type we considered,
with the level of sensitivity that is needed for
the study of effects that are truly at the Planck-scale.
A first key concern comes from the fact that the type
of interferometric setups we considered in Section~3
requires that the interferometer functions with beams at two
different frequencies, and the sensitivity estimate here reported
in Section~2 shows that in order to achieve Planck-scale sensitivities
the intereferometer should work with these two frequencies at
a level of accuracy that is comparable to the accuracy presently
achieved in dealing with a single frequency.
The modern interferometers that achieve these remarkable
accuracies rely on optimization of all experimental
devices to the frequency emitted by the laser.
It should be studied how significant a loss of accuracy would
result from working with both the laser frequency and a doubled frequency,
and how to compensate for this loss of accuracy by introducing new
techniques and new devices.

In particular, properties of some
elements of our interferometric setups
may be different for the two frequencies; for example,
the reflection time for a mirror may depend on the frequency.
One may attempt to study these properties in advance in
laboratory at the needed level of accuracy, and then
develope some appropriate
compensation techniques in the interferometric setups.
One can also exploit the differences between the two
possibilities of a terrestrial interferometer (LIGO/VIRGO-type)
which relies on a large ratio optical-length/arm-length (large
number of reflections) and of a gigantic space interferometer (LISA-type)
in which one does not use any reflections (since the arm's length
is already huge to begin with).
This difference might render LISA-type setups less sensitive to
problems due to lack of performance of mirrors used at two
different frequencies.

Another key challenge is posed by the fact that, at least
in the two setups considered in the previous section,
it appears to be necessary to implement controlled variations
of a macroscopic distance $L$ ({\it e.g.} length of one of the arms of the
interferometer).
For the setup of Subsection~3.1, with splitting of beams
both in configuration and in energy space, a key element
of the analysis relies on variations of a macroscopic
distance which are of the order of the wavelength of
the laser beam, which might be easily implementable
with appropriate use of active mirror control techniques already in use
in interferometry.
In the closing remarks of Subsection~3.1 we also pointed out
that, unless (as just mentioned) one manages to test in advance
to high accuracy some key
elements of the interferometric setup, such as a possible
dependence on wavelength of a mirror reflection time,
the quantum-grvaity analysis based on the setup of Subsection~3.1
may require repeating the experiment with a macroscopically different
choice of the length of one of the arms, keeping all other aspects
of the setup unchanged.
A realization of the strategy described in Subsection~3.1 may therefore
require both microscopic and macropic changes of the length of
one of the arms of the interferometer, in different stages of
the measurement procedure.
The setup of Subsection~3.2, with splitting of beams
only in energy space, relies from the onset on
repeating the experiment at
macroscopically different values of the length $L$ of the
single arm of that interferometric setup.

In the setup of Subsection~3.2 it appears however natural contemplate
possible ways to replace the actual macroscopic variations
of $L$ with some other strategy giving the sought effect.
It is in fact conceivable that in the setup of Subsection~3.2 one might
obtain a time-varying phase difference without actually varying $L$,
but rather using one of these possibilities:
(i) by introducing a
time dependence in the SHG-like
mechanism that generates the two different frequencies,
(ii) by using three identical SHGs at three positions where
alternatively one of the second and third ones is deactivated,
(iii) by introducing a medium along the path followed by the
light rays and implementing some form of electrical manipulation of
the optical properties of the medium).
The advantage of these possibilities (i),(ii),(iii) is that one
might conceivably implement them in such a way that there is no
actual macroscopic variation of the length $L$ (it would be changed
only ``effectively") and that the change of configuration
({\it e.g.}, from the configuration with the third SHG ``on" to one in
which the third SHG is ``off") might be implemented without actually
intervenining mechanically/manually on the interferometric setup,
by using for example remote electrical controls affecting the
properties of the third SHG of another medium.

\section{Closing remarks}
We have proposed that a strategy for the use
of laser-light interferometers in
tests of models of Planck-scale departures
from conventional Lorentz symmetry, which are attracting interest
in the literature also in relation with the emerging puzzle
of observations of ultra-high-energy cosmic rays.
A first-level simple-minded comparison between the magnitude
of the Planck-scale effect and the sensitivity of modern interferometers
was found (Sec.~2) to provide encouragement.
The key element for the relevant interferometric studies
is the capability to work
simultaneously with two beams of different wavelength.
This is certainly doable but with present technologies would imply
a (possibly severe) loss in overall sensitivity.
Since technical developments will be required (and their specific
nature might affect the structure of the intereferometric setup which
is eventually adopted), our discussion
focused on illustrating two examples (Sec.~3)
of interferometric setups that
would allow to convert the relevant Planck-scale effects into
signals for which interferometers are well suited.
The fact that the two setups considered here differ in several
significant aspects encourages us to think that, should technical
obstructions be encountered in the development of one such setup,
it should be possible to eventually find alternative interferometric
setups in which the technical challenges can be handled.

As stressed in Sec.~4, an actual realization of our proposal
may require a relatively long time. It appears however that
present outlook of our proposal should be viewed at least
from the perspective that was adopted when the first
ideas on gravity-wave intereferometric studies appeared in the
literature of the 1960s and 1970s.
The first preliminary analysis reported here appears to suggest that
the objective is achievable, and we hope that this may provide
encouragement for future studies.

\section*{Acknowledgement}

We are very grateful to A.~R\"{u}diger for a careful critical reading of
a first draft of the manuscript, which allowed us to
update our sensitivity estimates and also improved our understanding
of the technical challenges here discussed in Section~4.
G.A.C.~also acknowledges  helpful discussions with  L.~Gammaitoni,
B.T.~Huffman, R.~Onofri, and F.~Ricci.
C.L.~acknowledges fruitful discussions with  U.~Fr\"{o}hlich,
A.~Ostendorf, B.~Roth, and S.~Schiller.

\end{document}